\documentstyle[12pt]{article}
\renewcommand{\theequation}{\arabic{section}.\arabic{equation}}
\def\begine{\begin{eqnarray}}
\def\ende{\end{eqnarray}}

\begin{document}

\baselineskip=16pt

\begin{flushright}
SU-4240-586\\
UCI-TR 94-38\\
September 1994\\
\end{flushright}
\vskip1.0cm
{\bf {
\centerline {\bf\Large EXCITED HEAVY BARYONS}
\vskip.5cm
\centerline {\bf\Large IN THE BOUND STATE PICTURE}
}}
\vskip.5cm
\centerline {\large Joseph Schechter}
\vskip.3cm
\centerline {\it Physics Department, Syracuse University,}
\centerline {\it Syracuse, NY 13244-1130}
\vskip.3cm
\centerline{ and}
\vskip.3cm
\centerline{\large Anand Subbaraman}
\vskip.3cm
\centerline{\it Physics Department, University of California,}
\centerline{\it Irvine, CA 92717. }
\vskip1.5cm
\normalsize
\centerline {\large Abstract}
\vskip.3cm

The orbitally excited heavy quark baryons are studied in the Callan Klebanov
bound state model with heavy spin symmetry. First, a compact description
of the large $N_c$, infinite heavy quark mass bound state wavefunctions
and the collective quantization is given.  In order to study the kinematical
corrections due to finite masses we motivate an approximate Schrodinger-like
equation for the bound state. The effective potential in this equation is
compared with the quadratic approximation (spherical harmonic oscillator)
to it. This oscillator approximation is seen to be not very accurate.
It is noted that the present experimental information
cannot be even qualitatively understood with the usual light sector
chiral Lagrangian containing only light pseudoscalar mesons. The addition
of light vector mesons helps to overcome this problem.

\newpage

\baselineskip=18pt

\section{Introduction}
\setcounter{equation}{0}

The ``bound state'' picture \cite{1,2}, in which a baryon containing a
heavy quark is visualized as a bound state of a nucleon-as-Skyrme soliton
with a heavy meson, is a very appealing one. It has the interesting feature
that experimental information from the mesonic sector of the theory
(representing an approximation to QCD in the large $N_c$ limit) can be
used to predict the properties of the heavy baryons. Originally it was
applied \cite{1,2} to studying the ordinary hyperons, but a straightforward
extension was also made \cite{3} to the $c$ and $b$ baryons. After the
recognition of the importance of the Isgur-Wise heavy spin symmetry
\cite{4}, the study of the $c$ and $b$ baryons was pursued by seveal
groups \cite{5}-\cite{10}. The relatively large number of papers suggests
the richness and technical complexity of the approach.

The present paper was stimulated by experimental evidence \cite{11}
for two orbitally excited heavy $c-$baryons which may be interpreted as
the heavy spin multiplet $(\Lambda_c',\, \Lambda '^*_c )$ with
spin-parities $({1\over 2}^-, \, {3\over 2}^-)$. Some properties were
already studied in a bound state inspired framework \cite{12} based on
a potential for the spatial wavefunction of the form
\begine \label{sho}
V(r) = V_0 + {1\over 2}\kappa \, r^2 \, \, ,
\ende
where $V_0$ and $\kappa$ are constants. That treatment regarded $V_0$ and
$\kappa$ as arbitarary parameters to be fit. In the bound state approach
they are, however, computable in terms of the Skyrme profile functions
and the light meson - heavy meson coupling constants.  While the latter
especially are by no means precisely fixed it is easy to see that, in
models in which the only light mesons present are the pseudoscalars,
reasonable choices predict values for $V_0$ and $\kappa$ which give an
unbound or just barely bound ground state particle $\Lambda_c$. (The
binding is actually sizeable: $m(\Lambda_c)-m(D)-m(N) = -0.63$ GeV). Hence
there appears to be an important gap between the practical use of
(\ref{sho}) and the actual bound state calculations.

In order to investigate this problem we first formulate an approximate
Schrodinger like equation which should hold for finite heavy meson mass
$M$. The underlying equations that one gets, even for the ground state
\cite{9}, are not of the simple Schrodinger form, but comprise three
coupled differential equations. We therefore made a simple approximation
which should be good for large $M$ and which gives a Schrodinger-like equation
with a potential function which may be compared with (\ref{sho}). It turns
out that the quadratic approximation (\ref{sho}) is roughly reasonable for the
low-lying $b-$baryon energies but is significantly worse for the
low-lying $c-$baryon energies. The Skyrme potential gives more deeply bound
baryons than (\ref{sho}); however, the extra binding turns out to be
nowhere enough to solve the problem. The non-relativistic form of the
approximate Schrodinger equation enables us to easily make
very important ``two-body''
corrections corresponding to finite nucleon mass (which is infinite in the
large $N_c$ starting point) by introducing the reduced mass. It is
explained how this further reduces the ground state binding. The most
straightforward way to obtain reasonable values of the binding energy
seems to be to introduce light vector mesons \cite{7,8} in addition
to light pseudoscalars. These provide a great deal of extra binding
strength.   The requirement of
explaining the spectrum is noted to yield non-trivial constraints on the
light vector meson sector of the theory.

Another new aspect of this paper is the presentation of a somewhat
streamlined approach to the excited baryon wavefunctions. Following the
approach of refs. \cite{7,8} for the ground state we show how excited
state wavefunctions which are already diagonal (in contrast to those
of \cite{5} and \cite{10}) may be written down almost by inspection.
This is given in section 3 where the highly degenerate spectrum in
the $M \rightarrow \infty, \, N_c \rightarrow \infty$ limit is discussed,
including the effects of light vector mesons. The physical states of
the theory in this limit are recognized in section 4 after introducing
collective variables to describe the Skyrme ``tower''. This must be quantized
as a boson so the low-lying states look like the quark model ones
\cite{13} wherein a light diquark (belonging to the flavor $SU(3)$
representations $\bar{ {\mbox{\bf 3}}}$ or {\bf 6}) is rotating around a
heavy quark with effective orbital angular momentum $\ell_{eff}$. In turn,
$\ell_{eff}$ equals the ``light'' part of the ``grand spin'' of the heavy meson
in the background Skyrmion field.  Finally, the approximate Schrodinger
like equation for the physical kinematics situation is treated in
section 5 and conclusions are drawn.

\section{Notation}
\setcounter{equation}{0}

We will employ the same notations as in the earlier papers \cite{7} and
\cite{8}.  For the reader's convenience, a very brief reminder is
included here.

The total effective chiral Lagrangian is the sum of a ``light'' part
describing the three flavors $u,d,s$ and a ``heavy'' part describing the
heavy $(0^-, \, 1^-)$ meson multiplet $H$ and its interaction with the
light sector:
\begine
{\cal L}_{eff} = {\cal L}_{light} + {\cal L}_{heavy} \, \, .
\ende
The relevant light fields are the elements of the $3 \times 3$ matrix of
pseudoscalars $\phi$ and of the $3 \times 3$ matrix of vectors $\rho_\mu$.
Some objects which transform simply under the action of the chiral group
are
\begine
\xi =  e^{i\phi /F_\pi}\, , \qquad U = \xi^2 \, , \nonumber \\
F_{\mu\nu}(\rho) = \partial_\mu \rho_\nu - \partial_\nu \rho_\mu
 - i{\tilde g}[\rho_\mu, \rho_\nu]\, \, ,
\ende
where the pion decay constant $F_\pi \approx 0.132\, {\rm GeV}$ and
the vector meson coupling constant ${\tilde g} \approx 3.93$. References
on ${\cal L}_{light}$ may be traced from \cite{7,8}.

The heavy multiplet field combining the heavy pseudoscalar $P'$ and the
heavy vector $Q_\mu '$, both moving with a fixed 4-velocity $V_\mu$ is
given by
\begine\label{Hdef}
H = ({{1-i\gamma_\mu V_\mu}\over 2}) (i\gamma_5 P' + i\gamma_\nu Q_\nu ')
\, \, , \qquad {\bar H} = \gamma_4 H^\dagger \gamma_4 \, \, .
\ende
In this convention $H$ has the cannonical dimension one.  For
${\cal L}_{heavy}$ we take:
\begine \label{Lheavy}
{{{\cal L}_{heavy}}\over M} =&& i V_\mu {\rm Tr}[H(\partial_\mu - i \alpha
{\tilde g}\rho_\mu - i(1-\alpha) v_\mu) {\bar H}]
+id {\rm Tr}[H \gamma_\mu \gamma_5 p_\mu {\bar H}] \nonumber \\
&&+{{ic}\over {m_v}} {\rm Tr}[H \gamma_\mu \gamma_\nu F_{\mu\nu}(\rho)
{\bar H}] \, \, ,
\ende
where $M$ is the mass of the heavy meson, $m_v \approx 0.77\, {\rm GeV}$ is
the light vector meson mass and
\begine\label{vp}
v_\mu\, , \, p_\mu = {i\over 2}(\xi \partial_\mu \xi^\dagger \pm
  \xi^\dagger \partial_\mu \xi) \, \,  .
\ende
$d, \, c$ and $(\alpha {\tilde g})$ are respectively dimensionless coupling
constants for the $H-$light pseudoscalar, $H-$light vector magnetic type
and $H-$light vector gauge type interactions. It seems fair to say that
these coupling constants are not yet precisely fixed from experiment.
For definiteness we shall use the values
\begine\label{params}
d = 0.53 \, , \qquad c = 1.6 \, , \qquad \alpha = 1 \, \, .
\ende
The values for $d$ and $c$ are based on single pole fits
\cite{14,15} to the experimental
data \cite{16} on the $D \rightarrow K$ and
$D \rightarrow K^*$ semi-leptonic
transitions. The ratio $c/ d$ is consistent with that obtained \cite{17} using
a suitable notion of light vector meson dominance for the $D^* \rightarrow
D \gamma$ and $D^* \rightarrow D \pi$ branching ratios. It should be
remarked, however, that some authors have suggested \cite{18} a smaller value
of $d$, around 0.3. The value $\alpha = 1$ is based on light vector meson
dominance which \cite{17} seems reasonable.
Earlier \cite{8} we used a negative $\alpha$ based
on fitting the heavy baryon properties without using the excitation energy
constraint. The present paper contains a
reconsideration of that fit.

\section{Excited Baryon States at the Classical Level}
\setcounter{equation}{0}

In this section we first write the classical Skyrme solutions for the light
part of the action (which is taken to include the first three flavors).
Then we find the wavefunctions for which this ``baryon as soliton''
is bound to the heavy meson $H$. All the orbitally excited states, rather
than just the ground state, will be included. We shall work in the heavy
quark symmetry limit so that the radial wavefunctions reduce to their
delta-function limits. It turns out that the bound excited wavefunctions
are remarkably simple generalizations of the ground state one given in
\cite{7} and \cite{8} (see section 3 of \cite{8}, for example).

The usual hedgehog ans\"atz for the light pseudoscalars is
\begine \label{Fprofile}
\xi_c = \pmatrix{ {\rm {exp}}[i {\bf{\hat x}}\cdot\mbox{\boldmath $\tau$}
{{F(r)}\over 2}] & 0 \cr 0 & 1 } \, \, .
\ende
When we include the light vectors, we have similarly the classical
solutions
\begine
\rho_{\mu \, c}= \pmatrix{ {1\over {{\sqrt 2}}}(\omega_{\mu \, c} +
\tau^a \rho_{\mu \, c}^a ) & 0 \cr 0 & 1} \, \, ,
\ende
with
\begine\label{Gprofile}
\rho_{i \, c}^a = {1\over {{\sqrt 2}{\tilde g}r}} \epsilon_{ika}{\hat x}_k
G(r)\, , \hskip.3cm \rho_{0 \, c}^a = 0 \, , \hskip.3cm \omega_{i \, c}
= 0\, , \hskip.3cm \omega_{0 \, c} = \omega(r)\, \, .
\ende
The appropriate boundary conditions are
\begine\label{bc}
F(0)= -\pi \, ,\qquad G(0)=2 \, , \qquad \omega '(0)=0 \, , \nonumber \\
F(\infty)=G(\infty)=\omega(\infty)=0 \, .
\ende

Now, following the Callan-Klebanov approach, we want to find the
wavefunctions of a Schrodinger-like equation for ${\bar H}$ in the
classical background field above. Since the above ans\"atzae mix
isospin with orbital angular momentum, it is very convenient to make
a partial wave analysis in terms of the grand spin {\bf G},
\begine
{\bf G} = {\bf I} + {\bf L} + {\bf S} \, \, ,
\ende
where {\bf L} is the orbital angular momentum of the heavy meson field,
{\bf S} is its spin and {\bf I} its isospin. Due to the heavy spin
symmetry, the portion of {\bf S} from the heavy quark decouples from
the problem. Thus with the decomposition
\begine
 {\bf S} = {\bf S}' + {\bf S}'' \, \, ,
\ende
where {\bf S}$''$ is due to the heavy quark while {\bf S}$'$ is the
remainder, the truely relevant object is the ``light grand spin'' :
\begine \label{gspin}
 {\bf g} = {\bf I} + {\bf L} + {\bf S}'  \, .
\ende

We pointed out earlier \cite{7} that the {\it ground} state wavefunction
is characterized by $g=0$. Now we want to study the excited states.
Remember that in the heavy meson rest frame the $4 \times 4$ matrix
${\bar H}$ has non-vanishing elements only in the lower left $2 \times 2$
sub-block:
\begine
{\bar H} = \pmatrix{ 0 & 0 \cr {\bar H}^b_{lh} & 0 } \, .
\ende
Here the index $l$ represents the spin of the light degrees of freedom
within the heavy meson, the index $h$ stands for the heavy quark spin
and the index $b$ represents the isospin of the light degrees of freedom
within the heavy meson.  Following \cite{7} and \cite{8} we write the
general $g \not = 0 $ wavefunction as:
\begine \label{Hansatz}
{\bar H}^a_{lh}(g,g_3) = \left \{ \begin{array}{cl}
{{u(r)}\over {{\sqrt M}}} ({\bf {\hat x}}\cdot \mbox{{\boldmath $\tau$}})
_{ad}\, {\bar \psi}_{dl}(g,g_3)\, \chi_h \, , & a=1,2 \cr
 0 \, , & a=3 \, ,
 \end{array} \right .
\ende
wherein the radial function $u(r)$ satisfies $r^2 |u(r)|^2 \approx \delta(r)$
and $\chi_h$ is the heavy quark spinor. We will see how the presence of the
${\bf {\hat x}}\cdot \mbox{{\boldmath $\tau$}}$ factor simplifies the
calculations. The remaining factor ${\bar \psi}_{dl}$ is a kind of
generalized spherical harmonic with the appropriate covariance properties.
There is the standard three-fold ambiguity of which two vectors in
(\ref{gspin}) should be coupled together first. Here, coupling together
\begine \label{kspin}
{\bf K} = {\bf I} + {\bf S}'
\ende
leads to a nice further simplification. Then we write, for a
${\bar \psi}_{dl}$ with orbital angular momentum $r$ and ``K-spin'' $k$
in (\ref{kspin}) coupled to light grand spin $g$:
\begine \label{wavefcn}
{\bar \psi}_{dl}(g,g_3;r,k) = \sum_{r_3,k_3}C^{r \, k;g}_{r_3 \, k_3;g_3}
Y_{r,r_3} \xi_{dl}(k,k_3) \, ,
\ende
where $C$ stands for the Clebsch-Gordon coefficients, $Y$ stands for the
usual normalized spherical harmonics and $\xi_{dl}(k,k_3)$ are the
isospin-light spin wavefunctions:
\begine
\xi_{dl}(0,0) = {1\over {{\sqrt 2}}}(s^{\uparrow}_l i^{\downarrow}_d
- s^{\downarrow}_l i^{\uparrow}_d), \qquad \xi_{dl}(1,1) = s^{\uparrow}_l
i^{\uparrow}_d \,   , \nonumber \\
\xi_{dl}(1,0) = {1\over {{\sqrt 2}}}(s^{\uparrow}_l i^{\downarrow}_d
+ s^{\downarrow}_l i^{\uparrow}_d), \qquad \xi_{dl}(1,-1) = s^{\downarrow}_l
i^{\downarrow}_d \, \,  ,
\ende
$s$ and $i$ being the two component {\bf S}$'$ and {\bf I} spinors. Notice
that multiplying ${\bar \psi}_{dl}$ by  $({\bf {\hat x}}\cdot
\mbox{{\boldmath $\tau$}})_{ad}$, as in (\ref{Hansatz}), does not change its
$g-$spin.

So far, we have just written down possible wave functions allowed by
$g-$spin covariance; the question of which ones or linear combinations
actually correspond to bound states has not yet been addressed.  To
investigate this question we consider the matrix elements of the
``potential'' $V$ obtained from ${\cal L}_{heavy}$ in (\ref{Lheavy}).
(See (3.2) and (3.3) of \cite{7} for further details). The potential
conserves $g-$spin and parity so we need only its matrix elements between
wavefunctions $H$ and $H'$ of the same $g$ and parity:
\begine \label{pot}
&&{{-MdF'(0)}\over 2} \int d^3x\, (H')^{a'}_{hl'}\,({\bf {\hat x}}\cdot
\mbox{{\boldmath $\tau$}})_{a'b} \, \mbox{{\boldmath $\sigma$}}_{l'l}\cdot
\mbox{{\boldmath $\tau$}}_{bc}\, ({\bf {\hat x}}\cdot
\mbox{{\boldmath $\tau$}})_{ca} \, ({\bar H})^a_{lh} + \cdots
\nonumber \\
&&= {{dF'(0)}\over 2}\int d\Omega \, ({\bar \psi}')^*_{bl'} \,
\mbox{{\boldmath $\sigma$}}_{l'l}\cdot \mbox{{\boldmath $\tau$}}_{bc}
\, ({\bar \psi})_{al} \, (\chi '^\dagger \chi) + \cdots \, \, ,
\ende
wherein $d\Omega$ denotes the angular integration and the second line
follows from the first on the substitution of (\ref{Hansatz}) and the
observation that  the definition ${\bar H}= \gamma_4 H^\dagger \gamma_4$
for the 4-component matrix $H$ results in an extra minus sign for the
two component $H^a_{hl}$. The three dots in (\ref{pot}) stand for
contributions arising from the couplings of the light vectors to the
heavy meson; these do not change our conclusion and will be discussed
later.

Notice that all the $  {\bf {\hat x}}\cdot  \mbox{{\boldmath $\tau$}}$
factors have disappeared in the second line of (\ref{pot}). This means
that the effective potential operator for the wavefunction (\ref{wavefcn})
is simply:
\begine \label{simplepot}
{1\over 2}d F'(0) \, \mbox{{\boldmath $\sigma$}}\cdot
\mbox{{\boldmath $\tau$}} = {1\over 2}d F'(0)\left[ 2 {\bf  K}^2 - 3 \right]
\, \, ,
\ende
where we have used the fact that $\mbox{{\boldmath $\sigma$}}\cdot
\mbox{{\boldmath $\tau$}}$ is acting on wavefunctions of definite {\bf K}.
For $k=0$ we have the energy eigenvalue $-{3\over 2}d F'(0) \approx -$0.63
GeV, which indicates that the $k=0$ states are the bound ones. The
(in general) three $k=1$ states (corresponding to $r$ in (\ref{wavefcn})
taking on the values $g-1$, $g$, and $g+1$ ) have the energy eigenvalue
$+{1\over 2} g F'(0)$ and are unbound in this model. The wavefunctions
${\bar \psi}_{dl}$ in (\ref{wavefcn}) as well as the ${\bar H}$'s in
(\ref{Hansatz}) are already diagonal - a circumstance following from the
choice of $k$ as the intermediate label for constructing states of good
{\bf g}. It is not actually necessary to carry out the multiplication
by  $  {\bf {\hat x}}\cdot  \mbox{{\boldmath $\tau$}}$ in (\ref{wavefcn});
however, this is done in Appendix A. We shall refer to the ``$k-$value''
of a wavefunction as that of the factor ${\bar \psi}_{dl}$ in
(\ref{wavefcn}). As seen in Appendix A, this gets modified on multiplication
by $  {\bf {\hat x}}\cdot  \mbox{{\boldmath $\tau$}}$.

The bound state wavefunctions may be written in a very simple form. Since
they have $k=0$, (\ref{wavefcn}) simplifies to $Y_{gg_3}\, \xi_{dl}(0,0)$.
Furthermore $\xi_{dl}(0,0)= {1\over {{\sqrt 2}}}\epsilon_{dl}$, giving
finally for (\ref{Hansatz}):
\begine\label{boundwf}
{\bar H}^a_{lh}(g,g_3,s_3'') = \left \{ \begin{array}{cl}
{{u(r)}\over {\sqrt{ 2M}}} ({\bf {\hat x}}\cdot \mbox{{\boldmath $\tau$}})
_{ad}\, \epsilon_{dl}\, Y_{gg_3}\, \chi_h \, , & a=1,2 \cr
 0 \, , & a=3 \, .
 \end{array} \right .
\ende
The $g=0$ ground state wave function is seen, using $Y_{00}= 1/\sqrt{4 \pi}$,
to coincide with (3.7) of \cite{8}. In fact, the bound orbitally
excited wavefunctions are simply obtained by multiplying the ground
state one by ${\sqrt 4 \pi} Y_{gg_3}$.

The parities of the bound state wavefunctions are given by the formula:
\begine\label{parity}
{\rm parity} = (-1)^g \, \, .
\ende
This follows most directly from the fact that a negative parity meson is
being bound in a linear combination of states with orbital angular momenta
$g-1$ and $g+1$ (see the first line of (A1)). Of the three (in general)
unbound states with given $g$, one has parity $(-1)^g$ and the other two
have parity $-(-1)^g$.

The light grand spin quantum number $g$ has yet further physical significance
for the bound state wavefunctions. Note that in the extreme limit in which
we are presently working all the bound states are degenerate in energy.
This degeneracy will be broken if one allows for finite masses. We then
expect a centrifugal contribution to the energy of the form
\begine \label{centri}
{1\over {2Mr^2}} \ell_{eff}(\ell_{eff} + 1) \, \, ,
\ende
which would raise the energies for orbital excitations. Callan and Klebanov
pointed out in their original paper \cite{1} that $\ell_{eff}$ differs
from the orbital angular momentum due to the isospin-angular momentum mixing
in the Skyrme ans\"atz. It turns out that, in fact,
\begine
\ell_{eff} = g \, \, .
\ende
To see this, note that the centrifugal energy operator is approximately at
small $r$ given by
\begine\label{centriop}
{1\over {2Mr^2}} ({\bf L}^2 + 2 + 4 {\bf I}\cdot {\bf L}) \, ,
\ende
as is mentioned in \cite{1} and as also emerges in the approximation for
finite $M$ which perserves the heavy quark spin multiplets (to be discussed
in Section 5 of this paper). With
$\mbox{\boldmath{ $\lambda$}} = {\bf L} + {\bf I}$, (\ref{centriop}) may be
rewritten as $(2\,\mbox{\boldmath{$\lambda$}}^2 - {\bf L}^2 +
{1\over 2})/(2Mr^2)$.
Now the bound state wavefunction is a linear combination of $\ell = g-1$
and $\ell = g+1$ pieces. Considering a basis in which
$\mbox{\boldmath{$\lambda$}}$
is diagonal, we note that the $\ell = g-1$ piece can only couple to light
grand spin $g$ for the choice $\lambda = g-{1\over 2}$. Then the centrifugal
energy becomes
\begine
{1\over {2Mr^2}}\left[ 2(g-{1\over 2})(g+{1\over 2})- (g-1)g + {1\over 2}
\right] = {1\over {2Mr^2}}g(g+1)\, .
\ende
Similarly the $\ell = g+1$ piece requires $\lambda = g+{1\over 2}$, which
leads to the same result.

To end this section we give the form of the potential operator when the
effects of light vector mesons are included. Then (\ref{simplepot})
should be replaced by:
\begine
\mbox{{\boldmath $\sigma$}}\cdot \mbox{{\boldmath $\tau$}}
\left[ {1\over 2}d F'(0) - {c \over {m_v{\tilde g}}}G''(0) \right]
+ {\bf 1}\left[ - {{\alpha {\tilde g}}\over {{\sqrt 2}}} \omega (0)\right]
\, \, .
\ende
In this formula $d$, $c$ and $\alpha$ are the heavy meson$-$light meson
coupling constants defined in (\ref{Lheavy}), while $F(r)$, $G(r)$ and
$\omega (r)$ are the pseudoscalar, $\rho$ meson and $\omega$ meson
soliton profile functions defined in (\ref{Fprofile})$-$(\ref{bc}). Notice
that the $\rho$ meson piece has the same
$\mbox{{\boldmath $\sigma$}}\cdot \mbox{{\boldmath $\tau$}}$ factor
as the pseudoscalar piece while the $\omega$ meson piece has simply a
{\bf 1} in the light spin$-$isospin spaces. This shows that the
wavefunctions of (\ref{Hansatz}) and (\ref{wavefcn}) are still diagonal.
The eigenvalues for the $k=0$ states and for the $k=1$ states are
now read off to be:
\begine\label{vectorpot}
 V(k=0)&=& -{3\over 2} d F'(0)+ {{3c} \over {m_v{\tilde g}}}G''(0) -
{{\alpha {\tilde g}}\over {{\sqrt 2}}} \omega (0) \, , \nonumber \\
V(k=1) &=& {1\over 2}d F'(0) -{c \over {m_v{\tilde g}}}G''(0) -
{{\alpha {\tilde g}}\over {{\sqrt 2}}} \omega (0) \, .
\ende
The quantitites $F'(0)$, $G''(0)$ and $\omega (0)$ are obtained by solving
the coupled differential equations which arise by minimizing the static
energy of ${\cal L}_{light}$ (describing the ordinary baryons). This
yields \cite{19} for a typical best fit to baryon and meson masses:
\begine \label{slopes}
F'(0)= 0.795 \, {\rm GeV}, \qquad G''(0)= -0.390 \, {\rm GeV},
\qquad \omega (0)= -0.094 \, {\rm GeV} \, \, .
\ende

It should be remarked \cite{20}
that the sign of $\omega (0)$ is linked to the sign
of the $\rho \omega \phi$ coupling constant; the sign has been chosen
\cite{19,20} to
yield a best fit to light baryon electromagnetic form factors. To be on
the conservative side one might want to consider the possibility of
reversing this sign. Now, substituting (\ref{params}) and (\ref{slopes}) into
(\ref{vectorpot}) yields
\begine\label{vectorpotnos}
V(k=0)&=& -0.63 - 0.62 + 0.26\,(-0.26) = -0.99 \,(-1.51)\,
{\rm GeV}, \nonumber \\
V(k=1)&=& +0.31 +0.21 + 0.26\,(-0.26) = +0.78 \,(0.26)\, {\rm GeV},
\ende
wherein the order of the terms in (\ref{vectorpot}) has been retained.
The numbers in parentheses correspond to reversing the sign of $\omega (r)$
in \ref{slopes}. We
may observe that the $\rho$ meson contribution ($G''(0)$ terms) strengthens
the attraction in the bound channel and also increases the repulsion in
the unbound channel. The $\omega$ term is not dominant but may have
significant effects. The binding in the
present approximation seems somewhat too large. However, finite $M$
effects will provide a substantial reduction.

\section{Physical States of the Model}
\setcounter{equation}{0}

The states of definite angular momentum and isospin emerge, in the
soliton approach, after collective quantization. The collective angle-type
variable $A(t)$ \cite{21,22} is introduced on the light fields as
\begine \label{rotxi}
\xi({\bf x},t)= A(t)\xi_c({\bf x})A^\dagger (t), \qquad
\rho_\mu({\bf x},t) = A(t)\rho_{\mu\,c}({\bf x})A^\dagger (t)\, \, ,
\ende
where $\xi_c({\bf x})$ and $\rho_{\mu\,c}({\bf x})$ are given in
(\ref{Fprofile})-(\ref{Gprofile}) and generalized angular velocities
$\Omega_k$ are defined by
\begine
A^\dagger {\dot A} = {i\over 2} \sum_{k=1}^8 \lambda_k \, \Omega_k \, ,
\ende
the $\lambda_k$ being $SU(3)$ Gell-Mann matrices. All the heavy meson states,
rather than just the ground state, may be included by introducing a
Fock representation for the field ${\bar H}$ as
\begine\label{fock}
{\bar H}({\bf x},t) = \sum_n {\bar H}_n({\bf x})\,e^{iE_n t} \, a_n^\dagger
\, \, ,
\ende
wherein the stationary states $n=\{g,g_3;r,k,s_3''\}$ are given in
(\ref{Hansatz}) and (\ref{wavefcn}) (the unbound states are also included)
while $a_n^\dagger$ is the creation operator for the
state $n$ with energy eigenvalue $E_n$.  The ``light part'' of
${\bar H}$ is subjected to a collective rotation with the replacement
\begine \label{rotH}
{\bar H}^a_{hl}({\bf x},t) \rightarrow A_{ab}(t){\bar H}^b_{hl}({\bf x},t)\, .
\ende

Now the collective Lagrangian, $L_{coll}$, is obtained by substituting
(\ref{rotxi}) and (\ref{rotH}) into $\int d^3 x ({\cal L}_{light}
+ {\cal L}_{heavy})$ and carrying out the spatial integration. The final
physical states are recognized after the quantization of  $L_{coll}$.
There are two ways in which this $L_{coll}$ differs from that of the
usual $SU(3)$ Skyrme model \cite{22}. The first is that instead of a term
$-{{{\sqrt 3}}\over 2} \, \Omega_8$ there is now (in the one heavy quark
subspace) a term $-{{{\sqrt 3}}\over 3} \, \Omega_8$. As discussed in
section 4 of \cite{8}, the physical significance of this fact is that there
now exists a constraint on the quantum states which requires the
``Skyrmion rotator'' to transform as an irreducible representation
$\{\mu \}$ of $SU(3)$ which contains a state with hypercharge $Y=2/3$.
This particular state must necessarily  have integer isospin. Furthermore
its spin, denoted ${\bf J}_s$, must equal its isospin. The two lowest
Skyrmion multiplets (higher ones are probably model artifacts) are
\begine \label{su3reps}
(\mu = {\bar {\bf 3}}, & J_s = 0) \nonumber \\
(\mu = {\bf 6}, & J_s = 1).
\ende
Note that the Skyrmion rotator behaves as a boson; at the collective level
it evidently describes a diquark state. This analysis is forced upon
us when we consider the case of three light flavors \cite{23} with the
attendant Wess-Zumino term. It agrees with the original picture \cite{1}
of ``spin-isospin'' transmutation; namely, at the collective level the
heavy meson field
\begine\label{transmute}
&&\mbox{ (i) loses its flavor quantum numbers} \nonumber \\
&&\mbox{ (ii) acquires a spin equal to {\bf G}}
\ende
The full baryon state is a product of the bosonic diquark rotator in
(\ref{su3reps}) and the fermionic bound state wavefunction corresponding
to (\ref{transmute}).

The second way in which $L_{coll}$ differs from that of the $SU(3)$
Skyrme model is the presence of the following term linear in
${\bf \Omega}$:
\begine\label{hypersplit}
&& M\int d^3 x \, \Gamma_j(\Omega)\, H^a_{hl}\, (\tau^j)_{ab}\,
{\bar H}^b_{lh}, \nonumber \\
&& \Gamma_j(\Omega) = ({1\over 2}- \alpha)\Omega_j - (1-\alpha)\,
{\hat {\bf x}}\cdot {\bf \Omega} \, {\hat x}_j \, \, ,
\ende
where ${\bar H}$ is taken from  (\ref{fock}) and $\Gamma_j(\Omega)$
is evaluated near the origin of {\bf x}-space. From the analysis of
\cite{1}, one might expect a term like (\ref{hypersplit}) to lead to
(hyperfine) splitting of the baryons belonging to a given heavy spin
multiplet. Fortunately, this does not happen as one may see by inspection
using the bound (negative energy) wavefunctions of (\ref{boundwf}). Then
(\ref{hypersplit}) contains a factor:
\begine
\epsilon_{d'l}\,({\hat {\bf x}}\cdot \mbox{{\boldmath $\tau$}})_{d'a}\,
(\tau^j)_{ab} \, ({\hat {\bf x}}\cdot \mbox{{\boldmath $\tau$}})_{bd}\,
\epsilon_{dl} = {\rm Tr}({\hat {\bf x}}\cdot \mbox{{\boldmath $\tau$}}\,
\tau^j \,{\hat {\bf x}}\cdot \mbox{{\boldmath $\tau$}}) = 0 \, .
\ende
Thus, in the heavy mass $M \rightarrow \infty$ limit, the bound
eigenstates of the collective Hamiltonian are simply products of the
wavefunctions (\ref{boundwf}) with the diquark Skyrme rotator wavefunctions.
Taking account of (\ref{transmute}\,ii), the total angular momentum
{\bf J} is given by
\begine
{\bf J} = {\bf J}_s + {\bf G} = {\bf J}_s + {\bf g} + {\bf S}'' \, ,
\ende
where, for convenience, the heavy quark spin {\bf S}$''$ has been made
explicit in the last step. The Skyrme rotator wavefunctions are \cite{24}:
\begine
\Psi_{rot}(\mu, YII_3,J_s M) = (-1)^{J-J_3} \sqrt{dim \mu}\;\; D^{
(\mu)*}_{Y,I,I_3;{2\over 3}, \, J_s,\,\,-M_s}(A) \, ,
\ende
where $D^{(\mu)}(A)$ is the representation matrix of $SU(3)$. Defining
the total light spin ${\bf j} = {\bf J} - {\bf S}''$, we write the
overall wavefunctions for the bound states as:
\begine\label{grandwf}
\Psi^*(\mu; Y,I,I_3;J_s,g,j;j_3,s''_3) = \sum_{M,g_3}C^{J_s,g;j}
_{M,g_3;j_3} \, \Psi_{rot}^*(\mu, YII_3,J_s M)\, {\bar H}^a_{lh}
(g,g_3,s''_3) \, \, ,
\ende
where ${\bar H}^a_{lh}$ is given in (\ref{boundwf}). Finally
{\bf j} and {\bf S}$''$ may be added using $C^{j,{1\over 2};J}_{j_3,s_3;J_3}$
to yield the heavy spin baryon multiplets having $J = j \pm {1\over 2}$.

All the states in (\ref{grandwf}) have, up to relatively small $O(1/N_c)$
corrections, the degenerate energy eigenvalues $V(k=0)$ in (\ref{vectorpot}).
The $O(1/N_c)$ corrections associated with the present collective
quantization splits the ${\bar {\bf 3}}$ and ${\bf 6}$ $SU(3)$
representation states according to \cite{8}:
\begine
M({\bf 6})- M({\bar {\bf 3}}) = {2\over 3}\left[ m(\Delta)-m(N)\right] \, ,
\ende
where $\Delta$ and $N$ stand for the $\Delta(1230)$ and nucleon masses.
Now let us ennumerate the physical states. We expect, as mentioned after
(\ref{centri}), that finite $M$ will split the huge degeneracy so that $g=0$
corresponds to the ground states, $g=1$ to the first excited states and
so on. For $g=0$ the Clebsch Gordon addition in (\ref{grandwf}) becomes
trivial and the discussion reduces to that given in section 4 of \cite{8}.
All the $g=0$ states have (see (\ref{parity})) positive parity. The
$SU(3)$ representation ${\bar {\bf 3}}$ with $j=J_s=0$ has the
content $\{ \Lambda_Q,\, \Xi_Q ({\bar {\bf 3}}) \}$. It has spin
$J={1\over 2}$ obtained by adding the heavy spin $s''={1\over 2}$ to
$j=0$.  The $SU(3)$ representation {\bf 6} has $j=J_s=1$. Adding
$s''$ to this yields a degenerate heavy spin multiplet with both
$J={1\over 2}$ and $J={3\over 2}$ members. The flavor content is denoted
$\{ \Sigma_Q,\, \Xi_Q({\bf 6}),\, \Omega_Q\}$ and
$\{ \Sigma_Q^*,\, \Xi_Q^*({\bf 6}),\, \Omega_Q^* \}$.

It is sufficient to consider the first orbitally excited states with
$g=1$ in order to see the general pattern. All the $g=1$ states have,
according to (\ref{parity}), negative parity. The $SU(3)$  ${\bar {\bf 3}}$
diquark rotator state with $J_s=0$ is coupled in (\ref{grandwf}) to
total light spin $j=g=1$. Combining this, in turn, with the heavy quark
spin {\bf S}$''$ yields a heavy spin multiplet with both $J={1\over 2}$
and $J={3\over 2}$ members. These are denoted as
$\{ \Lambda '_Q,\, \Xi '_Q ({\bar {\bf 3}}) \}$ and
$\{ \Lambda '^*_Q,\, \Xi '^*_Q ({\bar {\bf 3}}) \}$. The $SU(3)$ {\bf 6}
diquark rotator state with $J_s=1$ is coupled in (\ref{grandwf}) to total
light spin $j=0,1\,{\rm or}\, 2$. Combining each of these three with
{\bf S}$''$ yields three heavy spin multiplets with total spin contents
($J={1\over 2}$),  ($J={1\over 2}$, $J={3\over 2}$) or
($J={3\over 2}$, $J={5\over 2}$). For each of these there are  6
flavor states analogous to $\{ \Sigma_Q,\, \Xi_Q({\bf 6}),\, \Omega_Q\}$.
Notice that the zero isospin states denoted $\Lambda'_c$ and
$\Lambda'^*_c$ are possible candidates for recently discovered \cite{11}
resonances.

$SU(3)$ splittings in the $M\rightarrow \infty$ limit have been discussed
in section 5 of \cite{8} for the $g=0$ states; a similar analysis can
be given for the orbitally excited states.

The positive energy (unbound) states, corresponding to $k=1$ in
(\ref{wavefcn}), can be ennumerated in a similar way using the appropriate
${\bar H}^a_{lh}(g,g_3; r,k,s''_3)$ in (\ref{grandwf}). While these
states would presumably not be bound in the true theory, they can be
expected to play a role as virtual intermediate states for further
application of the formalism.

\section{Kinematic Corrections}
\setcounter{equation}{0}

The preceeding results hold in the large $N_c$ limit (where the nucleon
mass is formally infinite) and in the large $M$ limit (where the heavy meson
mass is formally infinite). In order to compare with experiment it is clearly
important to get an idea of the corrections to be expected in the real world.
A complete discussion of these effects would be enormously complicated and
beyond the scope of the present investigation. Thus we will content
ourselves with a somewhat schematic model which has the advantage of
simplicity and which is familiar enough to stimulate our intuition. The
most straightforward way to proceed is to start with the simplest
``ordinary field'' Lagrangian which reduces to the heavy field Lagrangian
(\ref{Lheavy}) in the $M \rightarrow \infty$ limit. Such an ordinary
field Lagrangian, constructed with a heavy pseudoscalar $SU(3)$ triplet
$P$ and a heavy vector $SU(3)$ triplet $Q_\mu$ was given in (3.25) of
\cite{14} in connection with our original discussion of (\ref{Lheavy}).
For additional simplicity we shall at first neglect the heavy meson
interactions
with the light vectors, although we expect from (\ref{vectorpot}) and
(\ref{vectorpotnos}) that they are actually important. Then the heavy
Lagrangian becomes:
\begine\label{Lordi}
{\cal L}_{heavy} = &&-{\cal D}_\mu P  {\cal D}_\mu {\bar P}- M^2 P{\bar P} -
{1\over 2}\left({\cal D}_\mu Q_\nu - {\cal D}_\nu Q_\mu\right)
\left({\cal D}_\mu {\bar Q}_\nu - {\cal D}_\nu {\bar Q}_\mu\right)
- M^{*2} Q_\mu {\bar Q}_\mu \nonumber \\
&&+ 2iMd \left(P p_\mu {\bar Q}_\mu - Q_\mu p_\mu
{\bar P}\right) - id'\epsilon_{\beta\alpha\rho\mu} \left( {\cal D}_\rho
Q_\alpha p_\mu {\bar Q}_\beta - Q_\alpha p_\mu {\cal D}_\rho {\bar Q}_\beta
\right)\, \, ,
\ende
where ${\cal D}_\mu {\bar P} = \partial_\mu {\bar P} - i v_\mu {\bar P}$,
etc. Note that $v_\mu$ and $p_\mu$ are defined in (\ref{vp}).
In writing (\ref{Lordi}) we have allowed for different $P$ and $Q_\mu$
masses $M$ and $M^*$ as well as different coupling constants $d$ and
$d'$. But for simplicity we shall further restrict $M=M^*$ and
$d=d'$; thus our model will treat only those corrections to the
heavy quark symmetry due to finite $M$. The same model (\ref{Lordi})
has been investigated by Oh et. al. \cite{9} for the ground state heavy
baryons and with choices of $d \not = d'$ and $M \not = M'$ made to fit
the heavy baryon masses with experiment. Our treatment will differ in a
number of ways. First we shall approximate the three coupled differential
equations which result by a simple Schrodinger-like equation. We shall
generalize the approximate equation to include excited states, and
compare with the more standard picture \cite{12} of a quadratic potential.
We will also give some discussion of corrections due to the finite
nucleon mass.

To begin, we look for the stationary ground state solutions of the equations
of motion (see Appendix B) which result from (\ref{Lordi}). We know what
this should look like in the heavy limit.  Remember that the field $P$ above
is related to the field $P'$ in (\ref{Hdef}) by
\begine\label{pp'}
P = e^{imV\cdot x}P' \, \, ,
\ende
and similarly for $Q_\mu$. The ground state $(g=0)$ wavefunction for
${\bar H}$ in (\ref{boundwf}) translates to
\begine\label{Pground}
{\bar P}'^b &&= {{-i}\over 2} {\rm Tr}(\gamma_5 {\bar H}) = {{i}\over 2}
{{u(r)}\over {{\sqrt {8\pi M}}}} ({\hat {\bf x}}\cdot
\mbox{{\boldmath $\tau$}})_{bd} \, \rho_d  \, \, , \nonumber \\
{\bar Q}'^b_j &&=  {{-i}\over 2} {\rm Tr}(\gamma_j {\bar H}) ={{1}\over 2}
{{u(r)}\over {{\sqrt {8\pi M}}}} \left[ {\hat x}_j \, \delta_{bd} -
i({\hat {\bf x}} \times \mbox{{\boldmath $\tau$}}_{bd})_j \right]\, \rho_d
\, \, ,
\ende
where $b$ is the 2-valued isospin index and $\rho_d = \epsilon_{de}\chi_e$.
Taking (\ref{pp'}) into account and allowing different radial dependences
for the different terms in (\ref{Pground}) suggests the ans\"atzae \cite{9}
for the ordinary (unprimed) field stationary solutions:
\begine\label{Pansatz}
{\bar P}^b &&= {{\phi(r)}\over {{\sqrt {4 \pi}}}}({\hat {\bf x}}\cdot
\mbox{{\boldmath $\tau$}})_{bd} \, \rho_d \, e^{i\omega t}\, ,
\nonumber \\
{\bar Q}^b_j &&= {1\over {{\sqrt {4\pi}}}}\left[ i{\hat x}_j \, \delta_{bd}\,
\psi_1(r) + {1\over {{\sqrt 2}}} ({\hat {\bf x}} \times
\mbox{{\boldmath $\tau$}}_{bd})_j \, \psi_2(r)\right] \rho_d \,
e^{i \omega t} \, \, ,
\ende
where $\omega$ is the relativistic energy. At the ordinary field level we
should interpret $\rho_d$ as the isospace spinor. With this parametrization
the $M\rightarrow \infty$ limit is expressed as
\begine\label{heavylim}
\phi = -\psi_1 = -{{\psi_2}\over {{\sqrt 2}}} \, .
\ende
Substituting (\ref{Pansatz}) into the ${\bar P}$ equation of motion given
in (B.1) yields the differential equation:
\begine\label{eom}
\phi '' + {2\over r}\phi ' &&+ \left[ \omega^2 - M^2 + {1\over {r^2}} \{
-2 + {1\over 2}(1-\cos F)(3+ \cos F)\} \right] \phi - MdF' \psi_1
\nonumber \\
&&+ {{{\sqrt 2}Md \sin F}\over r} \psi_2 = 0 \, \, ,
\ende
where a prime denotes differentiation with respect to the radial coordinate
$r$. The two other similar coupled equations which result from (B.2) are
given in (B.4) and (B.5). First let us check the consistency of the
$M\rightarrow \infty$ limit. Dropping all terms not proportional to $M$
and substituting (\ref{heavylim}) we easily see that each of (\ref{eom}),
(B.4) and (B.5) reduces to
\begine
\left[ (\omega^2 - M^2) + Md(F' - {{2 \sin F}\over r})\right] \phi (r)=0\, .
\ende
This equation has a solution where $\phi(r)$ is sharply peaked around the
origin and where
\begine\label{heavysol}
E_{eff} = -{1\over 2}d\left[ F' - {{2 \sin F}\over r}\right]_{r=0}
= -{3\over 2}dF'(0) \, .
\ende
Here we have defined
\begine\label{Eeff}
E_{eff} = {{\omega^2 - M^2}\over {2M}} \approx \omega - M \, .
\ende
(\ref{heavysol}) is seen to agree with the binding energy $V(k=0)$
in (\ref{vectorpot}), when the contributions from the light vectors
are neglected.

Now let us approximate (\ref{eom}) to achieve an easy form for convenient
further study. It seems very natural to isolate the effects of finite $M$
by retaining it in (\ref{eom}) while approximating $\psi_1$ and $\psi_2$
by their $M \rightarrow \infty$ forms from (\ref{heavylim}). Then we end
up with a completely standard looking Schrodinger equation for the radial
wavefunction. We set $\phi (r) = w(r)/r$ as usual, divide through by
$2M$ and finally add the ``centrifugal'' term (\ref{centri}) so as to
generalize the equation to include orbitally excited partial waves.
Then (\ref{eom}) becomes
\begine\label{heavyeom}
-{1\over {2M}} w''(r) + \left[ {{\ell_{eff}(\ell_{eff}+1)}\over {2Mr^2}}
+ V_{eff}(r) \right]w(r) = E_{eff} \, w(r)\, \, ,
\ende
wherein,
\begine\label{Veff}
V_{eff}(r) = {{-d}\over 2} \left(F' -{{2\sin F}\over r}\right)
+{1\over {Mr^2}}\left[ 1-{1\over 4}\left(1-\cos F\right)\left(3+\cos F
\right)\right] \, \, .
\ende
Of course, $w(r)$ has an implicit $\ell_{eff}$ label. Even though
(\ref{heavyeom}) looks exactly like a non-relativistic Schrodinger equation
it actually (noting (\ref{Eeff})) contains the energy in the
characterestic relativistic manner, $(\omega^2-M^2)$. It describes a meson
of mass $M$ in the potential field $V_{eff}(r)$ due to an infinitely
heavy (large $N_c$ limit) nucleon. Notice that the ${1\over M}$ piece in
$V_{eff}$ which has an overall ${1\over {r^2}}$ factor behaves as $r^2$
for small $r$; this term is in fact very small. There is no exact
analytic form for the Skyrme profile $F(r)$. To make the equation
self-contained we adopt the Atiyah-Manton approximation \cite{25}:
\begine\label{Atiyah}
F(r) \approx -\pi \left[ 1 - {r \over {(\lambda^2 + r^2)^{1/2}}}\right]\, ,
\ende
where the parameter choice $\lambda^2 = 15.61584\,{\rm GeV}^{-2}$
corresponds to $F'(0)=0.795\, {\rm GeV}$ from (\ref{slopes}).
Using (\ref{Atiyah}), the effective potential $V_{eff}(r)$ is graphed in
Fig. 1. Also shown is the quadratic approximation which leads to a
spherical harmonic oscillator potential:
\begine\label{VSHO}
V_{SHO}= -{3\over 2}dF'(0) + {1\over 2}\kappa r^2 \, ,
\ende
where $\kappa$ is given in (4.4) of \cite{7}. With the neglect of the
light vectors and the choice (\ref{slopes}), we have $\kappa = 0.1562\,
{\rm GeV}^3$. It is seen that the quadratic approximation is not a very
accurate representation away from small $r$. This region is especially
relevant for orbitally excited states. On the other hand we note that,
as expected, the $V_{eff}(r)$ which results from the Skyrme approach is
not confining. Thus it is probably not trustworthy for higher orbitally
excited states.

To avoid confusion we remind the reader that, according to the discussion
of section 4, there are many physical states which correspond to a given
$\ell_{eff}$. They all have parity $(-1)^{\ell_{eff}}$. The heavy
baryons which belong to the $SU(3)$ ${\bar {\bf 3}}$ representations
comprise a heavy spin multiplet with spin content $(\ell_{eff}-{1\over 2}),
\,  (\ell_{eff}+{1\over 2})$ [for $\ell_{eff}=0$ this collapses to just
${1\over 2}$]. The heavy baryons which belong to the $SU(3)$ {\bf 6}
representation form three heavy spin multiplets with spin contents
$[(\ell_{eff}-{3\over 2}), \,  (\ell_{eff}-{1\over 2})]$,
$[(\ell_{eff}-{1\over 2}), \,  (\ell_{eff}+{1\over 2})]$ and
$[(\ell_{eff}+{1\over 2}), \,  (\ell_{eff}+{3\over 2})]$ [for $\ell_{eff}=0$
this collapses to just $({1\over 2},\, {3\over 2})$].

Now let us consider the numerical solutions of (\ref{heavyeom})-
(\ref{Atiyah}). For definiteness we take the weighted average of the
pseudoscalar and vector meson masses to obtain $M$; this yields $M=1.94 \,
{\rm GeV}$ for the charmed meson mass and $M=5.314\, {\rm GeV}$ for the
$b-$meson mass. The wavefunctions $w(r)$ are taken to behave as
$r^{(\ell_{eff}+1)}$ at the origin. Displayed in Table 1 are the values of
$E_{eff}$ corresponding to the most deeply bound state in each of the
$\ell_{eff}=0,\,1,\,2$ channels for both the $c-$baryons and $b-$baryons.
For comparison we also show the results obtained from the spherical
harmonic oscillator (SHO) approximation (\ref{VSHO}). In that case we have
a well known analytic formula for the energy levels:
\begine
E_{eff}^{(N)} = -{3\over 2}dF'(0) + \sqrt{{\kappa \over M}}\left(
{3\over 2}+N \right)
\, \, ,
\ende
where the level $N$ corresponds here to
states with $\ell_{eff}=N,\, N-2, \, \cdots $. It is interesting to
note that the exact numerical ground state ($\ell_{eff}=0$) energy eigenvalues
are more negative than the approximate SHO ones. This can be physically
understood from Fig. 1, since the SHO potential is narrower and should
thus have more zero point quantum fluctuation energy. As we expect, the
SHO approximation is better for the $b-$baryon case than for the lighter
$c-$baryon case where the more spread out
wavefunctions sample more of the differing large $r$ regions.
Similarly, the SHO approximation is better for the ground state than
for the excited states.
In fact the $c-$baryon  $\ell_{eff}=1$ state is unbound in the
SHO approximation. The model also
predicts radially excited wavefunctions. For the $c-$baryon another
$\ell_{eff}=0$ level is found at $-0.051$ GeV. In the SHO approximation
this is expected to be degenerate with the unbound $\ell_{eff}=2$
level at $+0.363$ GeV. As $M \rightarrow \infty$ the SHO approximation
results and the Skyrme potential results get closer to each other. For
example, already at $M=15$ GeV the difference between the predicted values
of the ground state $E_{eff}$ differ only by 1.5\%.
However, the collapse to the
$M\rightarrow \infty$ value of $-0.63$ GeV is rather slow; even at
$M=100$ GeV, there is a 10\% difference. Note that the finite $M$ effects
are large {\em both} for the $c$ and $b$ baryons.

\begin{table}
\centerline{
\begin{tabular}{|c| c| c|}
\hline
$~~~~~\ell_{eff}~~~~~~$ & $E_{eff}$ from (\ref{heavyeom})&
SHO approximation \\
\hline
\hline
{}~  & $c-$baryons & ~  \\
\hline
0 & $-0.277$ & $-0.204$ \\
1 & $-0.113$ & $+0.079$ \\
2 & $-0.012$ & $+0.363$ \\
\hline
{}~ & $b-$baryons  & ~  \\
\hline
0 & $-0.403$ & $-0.373$ \\
1 & $-0.277$ & $-0.205$ \\
2 & $-0.169$ & $-0.031$ \\
\hline
\end{tabular}}
\caption{ $E_{eff}$ in GeV from (5.10) and from the
spherical harmonic oscillator (SHO) approximation.}
\end{table}

A crucial question, of course, is how well these results agree with
experiment. For this purpose it will be seen to be sufficient to neglect
the relatively small $1/N_c$ corrections (of order 0.1 GeV). Then the
data at present yields just three relevant numbers. First there are the
``binding energies''
\begine
\mbox{B.E.}_c &=& m(\Lambda_c)-m(N)-m(D) = -0.63\, {\rm GeV} \, ,\nonumber \\
\mbox{B.E.}_b &=& m(\Lambda_b)-m(N)-m(B) = -0.78\, {\rm GeV}\, .
\ende
In addition, if the recently discovered \cite{11} zero isospin heavy baryons
are identified with the $\Lambda'_c$ and $\Lambda'^*_c$ mentioned in
section 4, we have the excitation energy
\begine
\mbox{E.E.}_c = m(\Lambda'_c) - m(\Lambda_c) = 0.31\, {\rm GeV}\, .
\ende
It seems natural to identify the binding energy with $E_{eff}$. Then the
comparison with experiment is presented in Table 2. In column (A) the
$M \rightarrow \infty$ results are given. The agreement with the binding
energies is reasonably good but the excitation energies are predicted to
be zero. The results of taking finite $M$ corrections into account via
the numerical solution of (\ref{heavyeom})-(\ref{Atiyah}) are shown in
column (B). We note that the general trend of the experimental data is
reproduced in the sense that each prediction of the model is about half of
the experimental value. The experimentally greater binding for the $b$
baryons compared to the $c$ baryons is reproduced. This picture is very
suggestive since we have been, for simplicity, working in a model which
does not include the light vector mesons. Equations (\ref{vectorpot}) and
(\ref{vectorpotnos}) show that one may expect the binding to be greatly
strengthened by the addition of the light vectors. To test this roughly
one may use the SHO approximation including light vectors. Taking the
inputs from (\ref{params}) and (\ref{slopes}) we have, in addition to
$V(k=0)=-0.99\,(-1.51)\, {\rm GeV}$ from (\ref{vectorpotnos}),
$\kappa = 0.295 \,(0.375)\, {\rm GeV}^3$ from (4.4) of \cite{7}, resulting
in
\begine\label{VSHOvect}
V_{SHO}\, (\mbox{with light vectors}) = -0.99 \, (-1.51) + 0.295\,(0.375)
{{r^2}\over 2} \, \, ,
\ende
where the numbers in parentheses correspond to the reversed sign for
$\omega (r)$. (Of course it is equivalent, from the present standpoint,
to keep the sign of $\omega (r)$ and reverse the sign of $\alpha$.)

The predictions from this model are shown in column (C) and are seen to
be significantly closer to experiment. (The values in parentheses show
the sensitivity to changing the sign of $\omega (r)$.) It thus appears
that the Callan Klebanov approach including light vector mesons and
${1\over M}$ corrections can roughly explain the general features of the
presently observed heavy baryon mass spectrum. This picture is however
achieved in the $N_c \rightarrow \infty$ limit in which the nucleon
mass is formally infinite. From a naive kinematical point of view this
seems peculiar, although we may perhaps argue that the $N_c \rightarrow
\infty$ limit is often more accurate than it has a right to be.

\begin{table}
\centerline{
\begin{tabular}{|c|c||c|c|c|c|}
\hline
{}~ & Expt. & (A) & (B) & (C) & (D)\\
\hline
$\mbox{B.E.}_c$ & $-0.63$ & $-0.63$ & $-0.277$ & $-0.41$ ($-0.85$)
& $+0.31$ ($-0.36$)\\
\hline
$\mbox{B.E.}_b$ & $-0.78$ & $-0.63$ & $-0.403$ & $-0.64$ ($-1.11$)
& $-0.08$ ($-0.48$)\\
\hline
$\mbox{E.E.}_c$ & 0.31  &  0    &  0.164 & 0.39 (0.44)
& 0.68 (0.77) \\
\hline
$\mbox{E.E.}_b$ &  ?    &  0    &  0.126 & 0.24 (0.27)
& 0.61 (0.69)\\
\hline
\end{tabular}}
\caption{ Comparison with experiment in the large $N_c$ limit.
All quantities are in GeV. Column (A) gives the prediction in the
$M \rightarrow \infty$ limit, column (B) corresponds to the solutions
of (5.10)-(5.12). Column (C) corresponds to the SHO
approximation when light vectors are included. Column (D) further includes
$N_c$ subleading two-body corrections due to finite nucleon mass to the
approximation (C).}
\end{table}

It is clearly of interest to explore the ``two body'' corrections
corresponding to taking the nucleon mass to be its finite experimental
value. The most straightforward way to proceed is to note that
(\ref{heavyeom}) may be regarded as a non-relativistic Schrodinger
equation with $E_{eff}$ the binding energy. Then we replace $M$ in
(\ref{heavyeom}) with the reduced mass
$\mu = (M \, m(N))/(M + m(N))$ according
to the usual prescription. If this is applied to finding the ground state
$\ell_{eff}=0$ energy for the heavy charmed baryon (where $\mu = 0.633$
GeV) we find $E_{eff}=-0.084$ GeV rather than the value $-$0.277 GeV
listed in Table 1. This drastic reduction in binding strength can
be easily understood in the SHO approximation. There the classical binding
energy of $-$0.63 GeV is pushed up by the zero point energy
${3\over 2}\sqrt{{\kappa \over \mu}}$. However the replacement $M \rightarrow
\mu$ greatly increases the zero point energy.

To get a somewhat more realistic impression we use the SHO approximation
with inclusion of the light vectors (\ref{VSHOvect}). Then we obtain
for the binding and excitation energies the results in column (D) of
Table 2.
It is clear that choosing the solution in parentheses, corresponding to the
choice of positive $\omega (0)$ in (\ref{slopes}), yields a fairly
reasonable picture for the ground state energies. The binding energies are
somewhat too small in magnitude but we have seen already for the model
without light vectors (Table 1) that there is more binding for the
$c-$baryons with the Skyrme profile than is indicated by the SHO
approximation. Thus we expect some improvements when we consider, in the
future, the analog of (\ref{heavyeom}) with the inclusion of light
vectors in $V_{eff}(r)$. Similarly, we expect considerable reductions in
the predicted excitation energies.
More precise determinations of the coupling
constants $d,\, c$ and $\alpha$ is evidently also important. Furthermore,
relativistic two body effects may play a role.

It does not seem that an adequate description of the spectrum can be
obtained in a model with just the light pseudoscalars included and when
finite $m(N)$ is taken into account - the lowest lying charmed
baryon would be barely bound.

In this paper we have presented a streamlined formalism for investigating,
in the Callan Klebanov bound state picture, the orbitally excited baryons
containing a heavy quark. The effects of light vector mesons and strange
quarks were included. Wavefunctions which diagonalize the bound state
Hamiltonian in the $M \rightarrow \infty, \, \, N_c \rightarrow \infty$
limit were obtained in a transparent way. These correspond to binding
energies which are somewhat too large. Furthermore there is no splitting
between any excited states and the ground state. For the purpose of
investigating the more realistic case we developed an approximate
Schrodinger like equation to describe the finite $M$ situation, but
where there is no breaking of the heavy spin multiplets.  The significant
difference
between the potential function of this equation obtained from the Skyrme
profiles of the light pseudoscalars and the spherical harmonic oscillator
approximation to it was pointed out. This was more important for the
$c-$baryons and for the excited states. The SHO approximation may be
considered reasonable only in a rough qualitative sense.
 The general effect of going to finite $M$, but still
infinite $m(N)$, was to reduce the strength of the binding. It was seen
that the model including only light pseudoscalars was unable to give
sufficient binding. With light vector mesons, however, the binding was
substantially increased. At this level the excitation energy was predicted
to be somewhat too low. Next, the effect of finite $m(N)$ was
taken into account in a non-relativistic approximation. This further
decreased the binding strength and favored one variant of the model
with light vectors.

In subsequent work we plan to study the effects of using the Skyrme rather
than the SHO potential (with light vectors) on the spectrum as well
as on the wavefunctions (and Isgur-Wise functions). We would also like
to further study heavy spin and $SU(3)$ breaking in the present framework.

\vskip.3cm

{\bf \large{Appendix A}}
\vskip.3cm
\setcounter{equation}{0}
\renewcommand{\theequation}{A.\arabic{equation}}

Here we display the effect of multiplying the wavefunctions
${\bar \psi}_{dl}(g,g_3;r,k)$ (defined in (\ref{wavefcn})) by
$({\hat {\bf x}}\cdot \mbox{{\boldmath $\tau$}})_{ad}$, as required for the
overall wavefunction given in (\ref{Hansatz}). For simplicity we shall
restrict ourselves to the case $g_3=g$ and use the shorthand notation
${\bar \psi}_{dl}(g,g_3;r,k)=\Phi(r,k)$. Then
\begine
({\hat {\bf x}}\cdot \mbox{{\boldmath $\tau$}})\, \Phi(g,0)&=&
{{-1}\over {\sqrt{2g+1}}}\left[ {\sqrt g}\, \Phi (g-1,1) -\sqrt{g+1}\,
\Phi(g+1,1) \right] \, ,\nonumber \\
({\hat {\bf x}}\cdot \mbox{{\boldmath $\tau$}})\, \Phi(g,1)&=&
{{-1}\over {\sqrt{2g+1}}}\left[ \sqrt{g+1}\, \Phi (g-1,1) + {\sqrt g}\,
\Phi(g+1,1) \right] \, ,\nonumber \\
({\hat {\bf x}}\cdot \mbox{{\boldmath $\tau$}})\, \Phi(g-1,1)&=&
{{1}\over {\sqrt{2g+1}}}\left[ -\sqrt{g+1}\, \Phi(g,1) - \sqrt{g}\,
\Phi(g,0) \right] \, ,\nonumber \\
({\hat {\bf x}}\cdot \mbox{{\boldmath $\tau$}})\, \Phi(g+1,1)&=&
{{1}\over {\sqrt{2g+1}}}\left[ -{\sqrt g}\, \Phi(g,1) + \sqrt{g+1}\,
\Phi(g,0) \right] \, .
\ende
The wavefunctions in the first line, with $k=0$, are the bound
ones. The others, with $k=1$, are unbound. Note that the wavefunctions
on the first two lines both have $\mbox{parity}=(-1)^g$ and are manifestly
orthogonal to each other. Those on third and fourth lines have
 $\mbox{parity}=-(-1)^g$ and are also manifestly orthogonal. In the
special case when $g=0$ there are just two, rather than four, independent
states - $\Phi(1,1)$ with positive parity which is bound, and
$\Phi(0,0)$ with negative parity which is unbound.

\vskip.3cm

{\bf \large{Appendix B}}
\vskip.3cm
\setcounter{equation}{0}
\renewcommand{\theequation}{B.\arabic{equation}}

The equations of motion which follow cannonically from (\ref{Lordi}) are:
\begine\label{Peom}
-{\cal D}_\mu{\cal D}_\mu \,{\bar P} + M^2 {\bar P} - 2iMd\,p_\mu
\,{\bar Q}_\mu=0 \, ,
\ende
\begine\label{Qeom}
&&-{\cal D}_\mu \left( {\cal D}_\mu \,{\bar Q}_\nu -{\cal D}_\nu \,{\bar Q}_\mu
\right) +M^{*2} {\bar Q}_\nu + 2iMd\,p_\nu {\bar P} \nonumber \\
&& +2id' \, \epsilon_{\beta\nu\mu\sigma}\,p_\mu \, {\cal D}_\sigma \,
{\bar Q}_\beta = 0\, \, .
\ende
We choose to deal with ${\bar P}$ rather than $P$ since ${\bar P}$
transforms like a light quark (rather than antiquark) and the interesting
heavy meson dynamics is associated with its light constituents.
(\ref{Qeom}) does not involve time derivatives of the component ${\bar Q}_4$.
Hence ${\bar Q}_4$ is non-dynamical and may be eliminated in terms of the
other fields. Correct to subleading $1/M$ order,
\begine\label{Q4}
{\bar Q}_4 \approx -{1\over {M^{*2}}} {\cal D}_i \, \partial_4 {\bar Q}_i \, .
\ende
Making the ans\"atzae (\ref{Pansatz}) for the ground state wavefunction and
substituting into (\ref{Peom})-(\ref{Q4}) yields, in addition to (\ref{eom})
the two additional equations:
\begine
&&-\psi_1'' - {2\over r} \psi_1' + \left[M^{*2} -\omega^2 +{2\over {r^2}}
(1+\sin^4{F\over 2})\right]\, \psi_1 \nonumber \\
&&+{\sqrt 2}\left[ -{{2 \sin^2{F\over 2}}\over {r^2}} +{{F'\sin F}\over
{2r}} +  {{\omega d'\, \sin F}\over r} \right] \, \psi_2
+ MdF' \, \phi = 0 \, \, ,
\ende
\begine
&&-\psi_2'' - {2\over r} \psi_2' + \left[M^{*2} -\omega^2 +{2\over {r^2}}
(1+2\sin^4{F\over 2}-2\sin^2{F\over 2})- \omega d' F' \right]\, \psi_2
\nonumber \\
&&+{\sqrt 2}\left[ -{{2 \sin^2{F\over 2}}\over {r^2}} +{{F'\sin F}\over
{2r}} +  {{\omega d'\, \sin F}\over r} \right] \, \psi_1
-{{{\sqrt 2}Md\, \sin F}\over r} \, \phi = 0 \, \, .
\ende

\vskip.3cm
{\large {\bf Acknowledgements:}} We would like to thank K. Gupta, A. Momen,
S. Vaidya and H. Weigel for helpful discussions. This work was supported in
part by the US DOE Contract No. DE-FG-02-85ER40231 and and NSF Grant No.
PHY-9208386.

\vskip.3cm

{\bf Figure Caption:}\\

Figure 1: Graph of the effective potential (5.11) with (5.12) due to
light pseudoscalar mesons. For comparison the graph of the quadratic
approximation (5.13) is also shown.

\end{document}